\documentclass[%
 reprint,
superscriptaddress,
 amsmath,amssymb,
 pre,
]{revtex4-2}

\usepackage{graphicx}
\usepackage{dcolumn}
\usepackage{bm}

\usepackage{xcolor}

\newcommand{\nvec}[0]{{\bm n}}
\newcommand{\uvec}[0]{{\bm u}}

\begin{document}


\title{Efficient single-precision simulations of nematohydrodynamics}
\author{Guilherme N. C. Amaral}
\affiliation{Centro de Física Teórica e Computacional, Faculdade de Ciências, Universidade de Lisboa, 1749-016 Lisboa, Portugal.}
 \affiliation{Departamento de Física, Faculdade de Ciências, Universidade de Lisboa, P-1749-016 Lisboa, Portugal.}
\author{Mahmoud Sedahmed}
\affiliation{Independent researcher, Cairo 11528, Egypt.}
\author{Margarida M. Telo da Gama}%
  \affiliation{Centro de Física Teórica e Computacional, Faculdade de Ciências, Universidade de Lisboa, 1749-016 Lisboa, Portugal.}
 \affiliation{Departamento de Física, Faculdade de Ciências,
Universidade de Lisboa, P-1749-016 Lisboa, Portugal.}
\affiliation{International Institute for Sustainability with Knotted Chiral Meta Matter, Hiroshima University, Higashihiroshima 739-8511, Japan.}
\author{Rodrigo C. V. Coelho}
\email{rcvcoelho@fc.ul.pt}
  \affiliation{Centro Brasileiro de Pesquisas Físicas, Rio de Janeiro 22290-180, RJ, Brazil}
\affiliation{Centro de Física Teórica e Computacional, Faculdade de Ciências, Universidade de Lisboa, 1749-016 Lisboa, Portugal.}
 \affiliation{Departamento de Física, Faculdade de Ciências, Universidade de Lisboa, P-1749-016 Lisboa, Portugal.}


\date{\today}

\begin{abstract}
Simulations of nematohydrodynamics on graphics processing units (GPUs) are typically performed using double precision, which ensures accuracy but significantly increases computational cost. However, consumer-grade GPUs are optimized for single-precision calculations, making double-precision simulations inefficient on widely available hardware. In this work, we demonstrate that single-precision simulations can achieve the same accuracy as double-precision methods while delivering a 27-fold increase in computational speed. To achieve this, we introduce two key improvements: (i) the shifted distribution function in the lattice Boltzmann method, which mitigates precision loss at low velocities, and (ii) the use of larger time steps in the finite-difference solver, which reduces numerical errors and improves overall accuracy. We find that, unlike in double precision, accuracy in single-precision simulations follows a non-monotonic trend with respect to the finite-difference time step, revealing an optimal regime for precise computations. To illustrate the effectiveness of our approach, we simulate the dynamics of single and multiple skyrmionic tubes in Poiseuille flow. Our results confirm that optimized single-precision simulations enable fast and accurate modeling of complex nematohydrodynamic systems, making large-scale simulations feasible on standard gaming GPUs.
\end{abstract}

\maketitle


\makeatletter
\newcommand{\manuallabel}[2]{\def\@currentlabel{#2}\label{#1}}
\makeatother

\section{Introduction}

Liquid crystals are a fascinating state of matter that combine fluidity with orientational order, making them distinct from conventional liquids and solids~\cite{Chen2018ER}. Their unique properties have led to numerous industrial applications, such as display technologies, as well as fundamental research into topological structures like skyrmions~\cite{Smalyukh2010, Ackerman2014, Ackerman2017, Fukuda2011, Posnjak2016, Ackerman2017b, guo:2016, tai:2019,  Zhao2023, sohn2018, PhysRevResearch.5.033210}. Liquid crystals exist in several phases, each characterized by different degrees of positional and orientational order. The smectic phase exhibits additional positional order, with molecules forming layered structures. In the cholesteric phase, the director field twists in a helical fashion, leading to unique optical properties. Among the various phases, the nematic phase stands out due to its ability to flow while maintaining a degree of orientational order. We will focus on this phase in this work.

To describe the dynamics of nematic liquid crystals, many theoretical frameworks have been developed, including the Beris-Edwards and Ericksen-Leslie models~\cite{Ericksen1962, doi:10.1098/rspa.1968.0195, stewart2019static, beris1994thermodynamics}. An important difference between them lies in the treatment of the scalar order parameter, which can vary in the Beris-Edwards model but remains constant in the Ericksen-Leslie one. The latter has proven successful in describing complex flowing topological structures such as skyrmions and torons~\cite{coelho2024halltransportliquidcrystal, Coelho_2021, mi15111302, Amaral2025}. However, simulating such systems poses significant computational challenges due to the enormous difference in time scales between the director field dynamics and the flow field evolution, which typically differ by six orders of magnitude~\cite{Coelho_2021}.

Previous works have assumed that this difference in the time scales is infinite and considered the relaxation of the flow field for a static configuration of directors~\cite{Coelho_2021, PhysRevE.89.032508}. This might give reasonable results in the absence of strong gradients. However, it is not applicable, for instance, near the two defects that are formed in toron configurations close to the parallel plates as strong spurious currents would appear in those regions. 

In previous works, we significantly reduced this problem by using the same time step for both methods, which needs to be very small~\cite{coelho2024halltransportliquidcrystal, mi15111302, Amaral2025}. We used a hybrid numerical approach: the lattice Boltzmann method to solve the flow field and a finite-difference scheme to evolve the director field. The parallelization of the code for GPUs was crucial to run the simulations in feasible times.

Here, we present a significant improvement to this method, focusing on the use of single-precision calculations in gaming GPUs, which are both more affordable and widely available than scientific GPUs. Two key advancements are introduced: (i) a shifting technique in the distribution function of the lattice Boltzmann method, which enhances precision in scenarios involving small velocities (common in liquid crystal simulations); and (ii) the use of larger time steps in the finite difference scheme, effectively reducing computational costs. We observe that an optimal time step exists that maximizes accuracy while maintaining the efficiency of the single-precision calculations.
This improved methodology expands the accessibility of large-scale liquid crystal simulations, making accurate computations feasible on consumer-grade hardware.

This paper is organized as follows. In Sec.~\ref{method-sec}, we describe the theory and the proposed method used to simulate hydrodynamics of liquid crystals. In Sec.~\ref{results-sec}, we describe the results using as a test case the simulation of a 3D skyrmion in Poiseuille flow. The single precision calculation is compared with double precision calculations and the different time steps are discussed. Finally, in Sec.~\ref{conclusion-sec}, we summarize and conclude.

\section{Method}
\label{method-sec}

In this section, we describe the theory and numerical method used to simulate the liquid crystal. We start by summarizing the hydrodynamic equations that govern the system. Then, we describe the finite-differences scheme and the lattice Boltzmann method and finalize by explaining how the two methods are coupled.

\subsection{Hydrodynamic equations}

The dynamics of the liquid crystal (LC) director field are governed by the Ericksen-Leslie model~\cite{Ericksen1962, doi:10.1098/rspa.1968.0195, stewart2019static}. This framework couples two dynamical equations: one describing the material flow and another governing the director field. These equations are well-suited for characterizing the hydrodynamics of LCs in the nematic phase.

For the velocity field, we employ the Navier-Stokes equation together with the continuity equation:
\begin{eqnarray}
&&\rho \partial_t u_\alpha + \rho u_\beta \partial_\beta u_\alpha = \partial_\beta \left[ - p \delta_{\alpha\beta} + \sigma_{\alpha\beta}^{v} + \sigma_{\alpha\beta}^{e} \right] + \rho g_\alpha \label{NS-eq} \\
&& \partial_\alpha u_\alpha = 0\label{cont-eq},
\end{eqnarray}
where the viscous stress tensor is defined as:
\begin{eqnarray}
\sigma_{\alpha\beta}^{v} = && \alpha_1 n_\alpha n_\beta n_\mu n_\rho D_{\mu\rho} + \alpha_2 n_\beta N_\alpha + \alpha_3 n_\alpha N_\beta \nonumber \\
&&+ \alpha_4 D_{\alpha\beta} + \alpha_5 n_\beta n_\mu D_{\mu\alpha} + \alpha_6 n_\alpha n_\mu D_{\mu\beta} .
\end{eqnarray}
In these expressions, $\rho$ represents the fluid density, $g$ is the external acceleration which drives the fluid motion, $p$ is the pressure, $\uvec$ is the velocity field, $\nvec$ denotes the director field describing the molecular alignment direction, and $\alpha_n$'s are the Leslie viscosities.

The kinematic transport, which accounts for the effect of the macroscopic flow on the microscopic structure, is given by:
\begin{eqnarray} 
N_\beta = \partial_t n_\beta + u_\gamma \partial_\gamma n_\beta - W_{\beta \gamma} n_\gamma 
\end{eqnarray} 
while the shear rate and vorticity tensors are defined as:
\begin{eqnarray} 
D_{\alpha\mu} = \frac{1}{2}\left ( \partial_\alpha u_\mu + \partial_\mu u_\alpha \right), \: W_{\alpha\mu} = \frac{1}{2}\left ( \partial_\mu u_\alpha - \partial_\alpha u_\mu \right). 
\end{eqnarray}
The elastic stress tensor takes the form:
\begin{eqnarray} 
\sigma_{\alpha\beta}^{e} = -\partial_\alpha n_\gamma \frac{\delta E}{\delta (\partial_\beta n_\gamma)}, 
\end{eqnarray} 
where $E$ denotes the Frank-Oseen elastic free energy:
\begin{eqnarray}
E =&& \int dV \biggl ( \frac{K_{11}}{2} (\nabla \cdot \nvec)^2 + \frac{K_{22}}{2} \left ( \nvec\cdot [\nabla \times \nvec] + q_0 \right)^2 \\ && + \frac{K_{33}}{2}[ \nvec\times [\nabla \times\nvec]]^2 \biggr) .
\label{free-energy-eq}
\end{eqnarray}
where $K_{11}$, $K_{22}$, and $K_{33}$ are the Frank elastic constants (splay, twist and bend respectively), and $q_0=2\pi/P$ with $P$ representing the cholesteric pitch.

The second set of equations describes the evolution of the director field:
\begin{align} 
& \partial_t n_\mu = \frac{1}{\gamma_1} h_\mu - \frac{\gamma_2}{\gamma_1} n_\alpha D_{\alpha\mu} - u_\gamma \partial_\gamma n_\mu + W_{\mu\gamma}n_\gamma , 
\label{director-time-eq} 
\end{align}
where $\gamma_1=\alpha_3-\alpha_2$ is the rotational viscosity, determining the relaxation rate of the director, and $\gamma_2 =\alpha_3+\alpha_2$ is the torsion coefficient, which characterizes the contribution of velocity field gradients to the viscous torque. The ratio $\gamma_2/\gamma_1$ is the alignment parameter, with $\vert \gamma_2/\gamma_1 \vert>1$ corresponding to flow-aligning and $\vert \gamma_2/\gamma_1 \vert<1$ to flow-tumbling systems. The molecular field is expressed as:
\begin{eqnarray}
h_\mu = -\frac{\delta E}{\delta n_\mu}. 
\end{eqnarray}

The simulations started with the liquid at rest, with the directors predominantly aligned perpendicularly to the plates, except in the vicinity of the toron. The toron configuration was obtained by minimizing its free energy from an initial Ansatz based on Ref.\cite{Coelho_2021}. The material parameters were chosen to match those of MBBA at 22$^\circ$C\cite{PhysRevE.89.032508}, except for the absolute viscosity (or, equivalently, $\alpha_4$), which was doubled to ensure reasonable simulation times while keeping its magnitude comparable to the material value. 

The physical behaviour is primarily governed by the Ericksen number~\cite{Coelho_2021}:
\begin{align} 
Er\equiv\frac{\mu v_s P}{K}, 
\end{align} 
where $v_s$ is the characteristic velocity (the skyrmion velocity for instance), $P$ the cholesteric pitch, $\mu$ the absolute viscosity, and $K$ the average elastic constant.

\subsection{Finite-differences method}
\label{FD-sec}

The simulations are based on a hybrid numerical approach as will be described in Sec.~\ref{hybrid-sec}. The velocity field was solved using the lattice Boltzmann method~\cite{kruger2016lattice,succi2018lattice}, as outlined in the following subsection, with elastic and viscous stress tensors incorporated as force terms. Thus, the following term is calculated using finite-differences to be used in LBM:
\begin{eqnarray}
    &&F_\alpha^{LC} = \partial_\beta \left[  \alpha_1 n_\alpha n_\beta n_\mu n_\rho D_{\mu\rho} + \alpha_2 n_\beta N_\alpha + \alpha_3 n_\alpha N_\beta \right. \label{force-lc-eq} \\ && \left. + \alpha_4^\prime D_{\alpha\beta} + \alpha_5 n_\beta n_\mu D_{\mu\alpha} + \alpha_6 n_\alpha n_\mu D_{\mu\beta} -\partial_\alpha n_\gamma \frac{\delta E}{\delta (\partial_\beta n_\gamma)} \right]. \nonumber
\end{eqnarray}
Note that the term $\alpha_4^\prime D_{\alpha\beta}$ is already considered in LBM, but we also keep this in the force term in order to achieve larger viscosities without the limitations of the relaxation time $\tau$ in the SRT collision operator of LBM. We choose the same viscosity both in LBM and in FD such that $\alpha_4 = \alpha_4^\prime + \rho c_s^2(\tau-1/2)$. Spatial derivatives were computed using central moment differences:
\begin{align} 
\frac{d\mathcal{F }}{dx} = \frac{\mathcal{F }(x+\Delta x) - \mathcal{F } (x-\Delta x)}{2\Delta x} + \mathcal{O}(\Delta x^2), 
\end{align} 
where $\mathcal{F }$ is a generic function and $\Delta x$ is the spatial step. Gradients were calculated solely at fluid nodes while keeping solid node values fixed. On solid boundaries, including obstacles, infinite homeotropic anchoring and no-slip conditions were enforced using the bounce-back scheme~\cite{kruger2016lattice}.

The director field equation, Eq.\eqref{director-time-eq}, was solved using a finite difference predictor-corrector scheme~\cite{Grossmann2007, Vesely2001}. The basic algorithm is summarized as follows:
\begin{itemize}
    \item Initialize the director field $n_\alpha$ and the auxiliary quantity $n_\alpha^P$, which is used in the predictor step. Both can be initialized with the same values.
    \item Calculate the derivative $(\partial _t n_\alpha)_{old}$ of the director field using $n_\alpha(t)$ in Eq.~\eqref{director-time-eq}.
    \item Calculate the predictor: $n_\alpha^P = n_\alpha(t) + \Delta t_{FD}(\partial _t n_\alpha)_{old}$, where $\Delta t_{FD}$ is the time step of the finite-difference scheme.
    \item Calculate the time derivative $(\partial _t n_\alpha)_{P}$ again but using the predictor $n_\alpha^P$ in Eq.~\eqref{director-time-eq}.
    \item Calculate the averaged time derivative: $\partial _t n_\alpha = \frac{1}{2}[(\partial _t n_\alpha)_{old} + (\partial _t n_\alpha)_{P}]$.
    \item Calculate the corrector: $n_\alpha = n_\alpha(t) + \Delta t_{FD}\partial _t n_\alpha$.
\end{itemize}
The force given by Eq.~\eqref{force-lc-eq} is calculated at every $\Delta t_{FD}$ together with the predictor-corrector scheme as described in Sec.~\ref{hybrid-sec}.

\subsection{Lattice Boltzmann method}
\label{lbm-sec}

We obtain the velocity field by solving the discretized Lattice Boltzmann Equation (LBE) using the single relaxation time (SRT) collision approximation and Guo’s forcing scheme \cite{PhysRevE.65.046308, Krger2017} as follows:
\begin{eqnarray}
    &&f_i (\mathbf{x} + \mathbf{e}_i \Delta t, t + \Delta t) - f_i (\mathbf{x}, t) = -\frac{\Delta t}{\tau} (f_i (\mathbf{x}, t) \nonumber \\
    &&- f_i^{eq} (\mathbf{x}, t)) + \Delta t \left( 1 - \frac{1}{2 \tau} \right) S_i^F (\mathbf{x}, t),
\end{eqnarray}
where $f_i$ is the distribution function, $f_i^{eq}$ is the equilibrium distribution function, given by:
\begin{eqnarray}
    f_i^{eq} (\mathbf{x}) &&= w_i \rho(\mathbf{x}) \left[ 1 + \frac{\mathbf{e}_i \cdot \mathbf{u}(\mathbf{x})}{c_s^2} + \frac{(\mathbf{e}_i \cdot \mathbf{u}(\mathbf{x}))^2}{2 c_s^4} \right. \nonumber \\ && \left.- \frac{\mathbf{u}(\mathbf{x}) \cdot \mathbf{u}(\mathbf{x})}{2 c_s^2} \right],
    \label{feq-eq}
\end{eqnarray}
where $c_s$ is the lattice speed of sound. The parameters $\Delta x$, $\Delta t$, and the reference density $\rho_0$ are chosen as unity (lattice units). The lattice weights $w_i$ and discrete velocity vectors $e_i$ for the D3Q19 lattice arrangement are~\cite{Krger2017}: $c_s=1/\sqrt{3}$, $w(\vert \mathbf{e}_i \vert ^2 = 0 ) = 1/3$, $w(\vert \mathbf{e}_i \vert ^2 = 1 ) = 1/18$ and $w(\vert \mathbf{e}_i \vert ^2 = 2 ) = 1/36$.

The macroscopic fluid density is calculated as:
\begin{equation}
    \rho(\mathbf{x}) = \sum_i f_i(\mathbf{x})
\end{equation}
while the macroscopic velocity is given by:
\begin{equation}
    \mathbf{u}(\mathbf{x}) = \frac{1}{\rho(\mathbf{x})} \sum_i \mathbf{e}_i f_i(\mathbf{x}) + \frac{\Delta t}{2 \rho(\mathbf{x})} \mathbf{F}(\mathbf{x}).
\end{equation}
The relaxation time $\tau$ is related to the kinematic viscosity of the fluid ($\nu$): $\nu = c_s^2 (\tau - \frac{1}{2}) \Delta t$. The forcing term in the LBE equation is:
\begin{eqnarray}
    S_i^F(\mathbf{x}) &&= w_i \left[ \frac{(\mathbf{e}_i \cdot \mathbf{F}(\mathbf{x}))}{c_s^2} \right. \nonumber \\ && \left.+ \frac{(\mathbf{e}_i \mathbf{e}_i - c_s^2 \mathbf{I}) : (\mathbf{u}(\mathbf{x})\mathbf{F}(\mathbf{x}) + \mathbf{F}(\mathbf{x})\mathbf{u}(\mathbf{x}))}{2 c_s^4} \right].
\end{eqnarray}
The bounce-back boundary condition is used to model the no-slip boundary condition over solid nodes\cite{kruger2016lattice}. The idea behind the bounce-back boundary condition is to reflect the distribution functions hitting a solid node with the same magnitude but in the reverse direction. The exact location of the solid wall is assumed to be located between two lattice points\cite{kruger2016lattice}.

\begin{figure}[htb]
\includegraphics[width=0.45\textwidth]{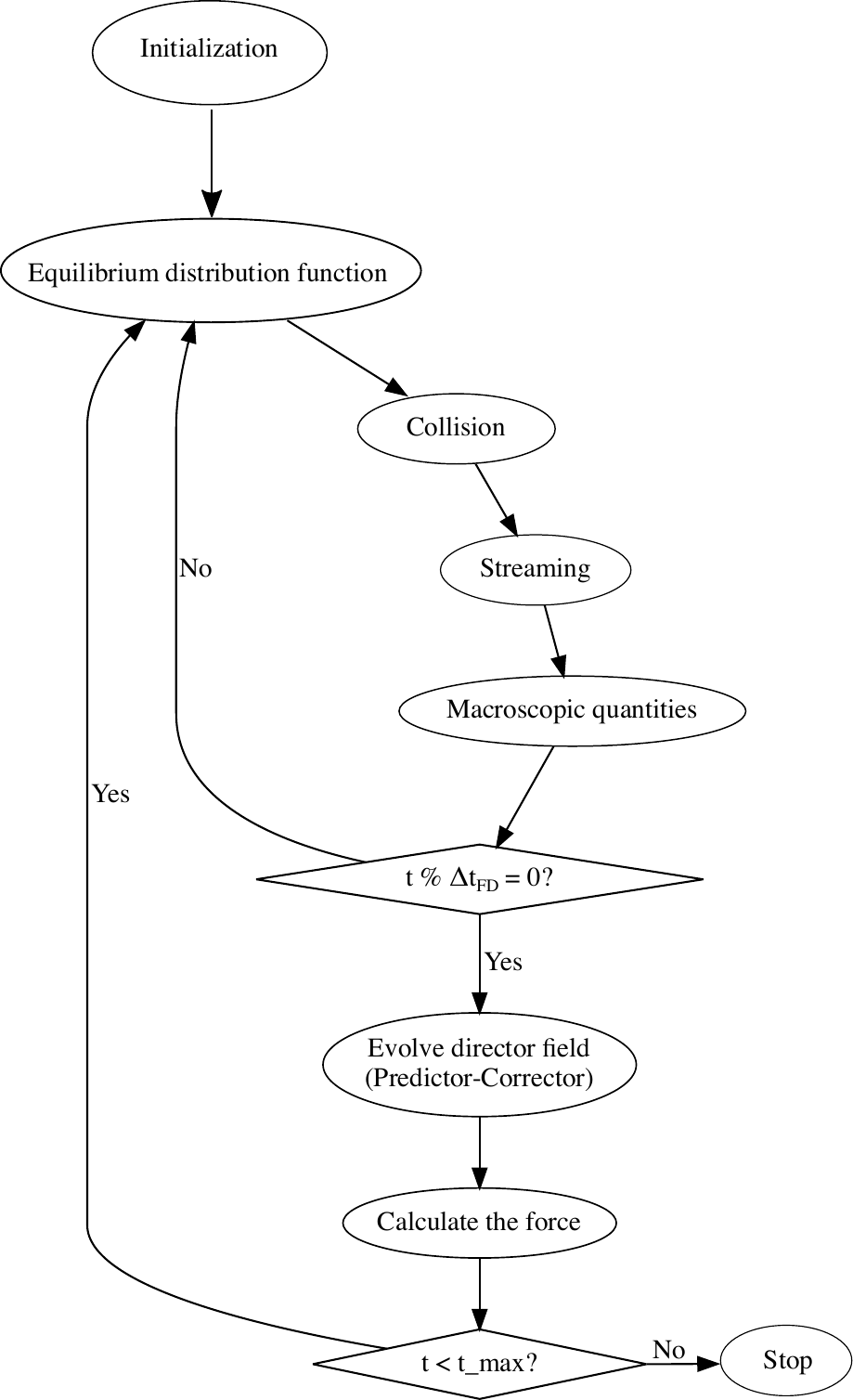}
\caption{Flowchart of the hybrid method.}
\label{fluxogram-fig}
\end{figure}

\begin{figure*}[htb]
\includegraphics[width=\textwidth]{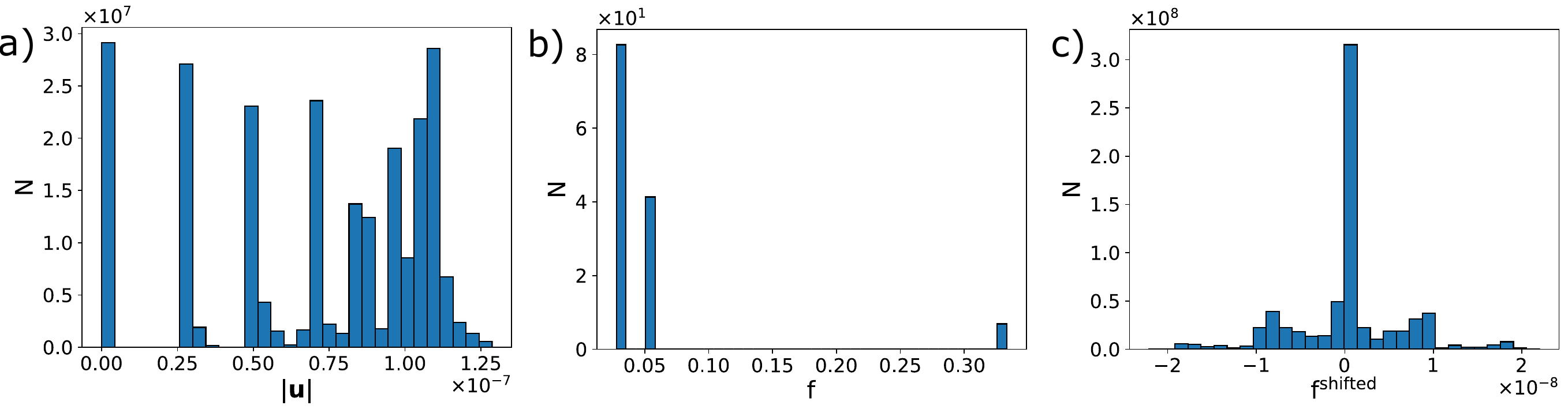}
\caption{Normalized histograms from a typical simulation using double precision, with $\Delta t _{FD}=1$, of (a) velocity field, (b) distribution function and (c) shifted distribution function.}
\label{histograms-fig}
\end{figure*}

We also used the distribution function shifting technique reported in \cite{computation4010011} to increase the computational accuracy of the model. The shifting technique is based on the observation that the values of the distribution function values are oriented around the weighting coefficients ($w_i$), which represent the zero velocity equilibrium distribution functions $f_i^{eq} (\rho_0 = 1, \mathbf{u}_0 = 0)$ based on Eq.\eqref{feq-eq}. Hence, instead of storing the full value of the distribution function, only the perturbation (shifted) value from the zero-velocity equilibrium value is stored and used in the computations. The shifted distribution function is defined as follows:
\begin{eqnarray}
    f_i^{\text{shifted}}(\mathbf{x}, t) = f_i(\mathbf{x}, t) - f_i^{eq}(\rho_0=1, \mathbf{u}_0=0).
    \label{shifted-feq}
\end{eqnarray}
The macroscopic parameters are obtained as:
\begin{eqnarray}
    &&\rho(\mathbf{x})-1 = \sum_i f_i^{\text{shifted}}(\mathbf{x}, t),\label{shifted-rho}\\
    &&\mathbf{u}(\mathbf{x}) = \frac{1}{\rho(\mathbf{x})} \sum_i \mathbf{e}_i f_i^{\text{shifted}}(\mathbf{x}, t) + \frac{\Delta t}{2 \rho(\mathbf{x})} \mathbf{F}(\mathbf{x}). \label{shifeted-u}
\end{eqnarray}
Moreover, the shifted equilibrium distribution function is given by:
\begin{eqnarray}
    f_i^{eq, \text{shifted}}(\mathbf{x}) &&= w_i \rho(\mathbf{x}) \left[ \frac{\mathbf{e}_i \cdot \mathbf{u}(\mathbf{x})}{c_s^2} + \frac{(\mathbf{e}_i \cdot \mathbf{u}(\mathbf{x}))^2}{2 c_s^4} \right. \nonumber \\ && \left. - \frac{\mathbf{u}(\mathbf{x}) \cdot \mathbf{u}(\mathbf{x})}{2 c_s^2} \right] + (\rho(\mathbf{x}) - 1) w_i.
    \label{shifted-feq}
\end{eqnarray}
The shifted LBE takes the form:
\begin{eqnarray}
    && f_i^{\text{shifted}}(\mathbf{x} + \mathbf{e}_i \Delta t, t + \Delta t) - f_i^{\text{shifted}}(\mathbf{x}, t)   \nonumber \\ && = -\frac{\Delta t}{\tau} (f_i^{\text{shifted}}(\mathbf{x}, t) - f_i^{eq, \text{shifted}}(\mathbf{x}, t)) \nonumber \\ && + \Delta t \left(1 - \frac{1}{2 \tau}\right) S_i^F(\mathbf{x}, t).
    \label{shifted-lbe}
\end{eqnarray}
where it can be observed that only the shifted distribution functions have to be stored to solve the LBE and the macroscopic parameters are easily obtained.
The shifting technique is critical to obtain proper simulation results using single-precision accuracy. For a standard single precision 32-bit IEEE 745 floating point data type, 4 bytes (32 bits) of memory are allocated to store the value and it provides around 7 decimal digits of precision. On the other hand, the double precision floating point data type requires 8 bytes (64 bits) of memory, and it provides around 15 decimal digits of precision. Hence, using single precision accuracy results in having fewer bits of memory to store the differences between the computed values, with the relative error being roughly ($10^{-7}$ ). For instance, Eq.~\eqref{shifeted-u} is used to determine the flow velocity components using the difference between the distribution functions in the different directions in addition to the value of the external force component. The small differences between the terms in Eq.~\eqref{shifeted-u} become critical to the accuracy and stability of the simulation. 
The values of the distribution function themselves in the model are typically around the values of the lattice weights ($w_i$), as shown in Fig.~\ref{histograms-fig}b. Moreover, the equilibrium distribution functions at zero velocity $f_i^{eq} (\rho_0=1,
\mathbf{u}_0=0)$ take the values of the lattice weights as well. Hence, subtracting $f_i^eq (\rho_0=1,\mathbf{u}_0=0)$ from $f_i (x,t)$ shifts the values of the distribution function to around zero instead of the lattice weights ($w_i$), as shown in Fig.~\ref{histograms-fig}c. Therefore, when the differences between shifted distribution functions are evaluated, the absolute error of the single precision accuracy (i.e., $10^{-7}$ times the largest numerical value being evaluated) was reduced as the evaluated numbers were reduced. 
Finally, the shifting technique is implemented by first evaluating the right-hand side of Eq.~\eqref{shifted-rho}, which provides the value of ($\rho(x)-1$) that can be substituted directly in Eq.~\eqref{shifted-feq}. Then, the LBE is solved via the typical collision and streaming algorithm of the LBM.

\begin{figure*}[htb]
\includegraphics[width=\textwidth]{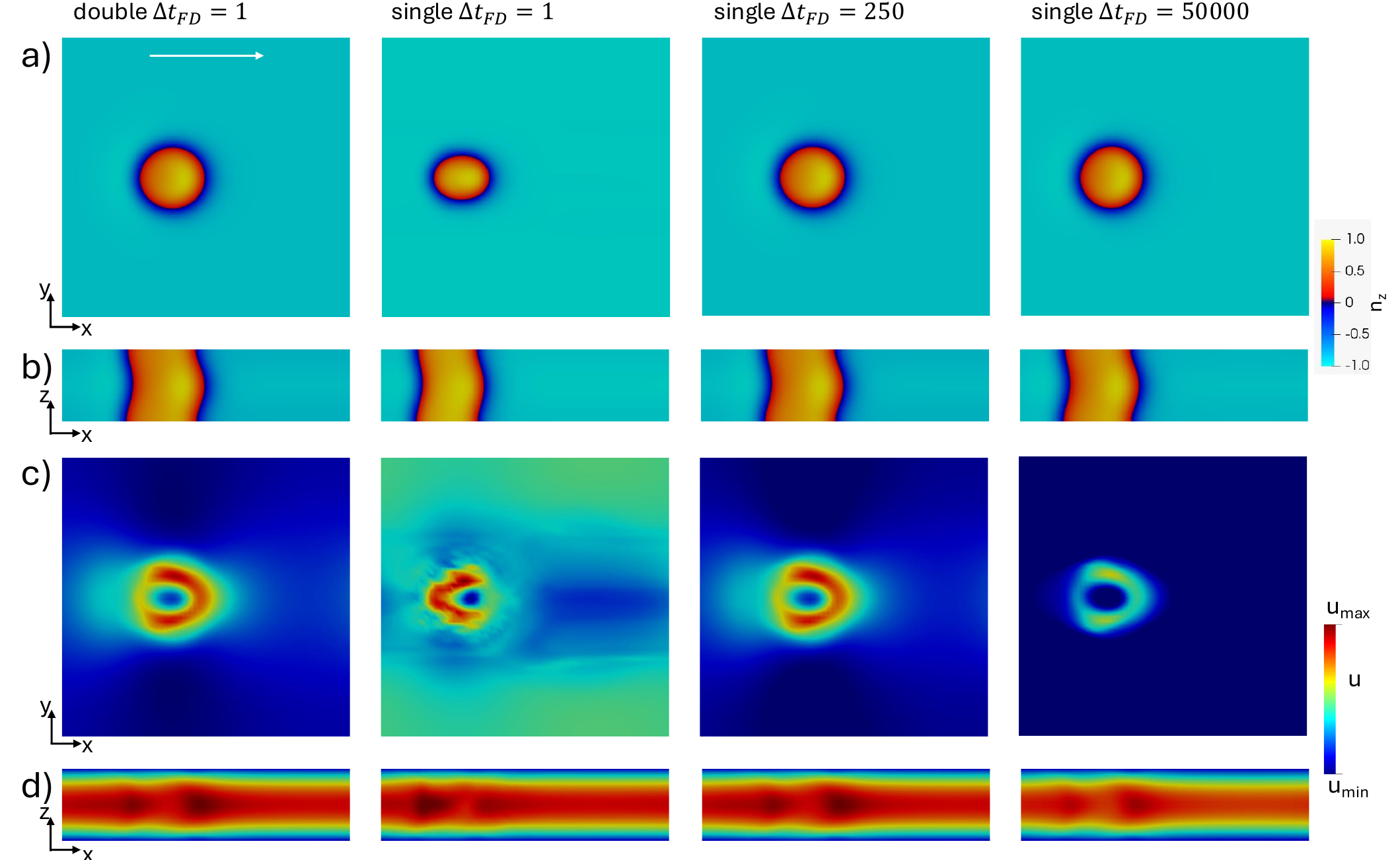}
\caption{Screenshots of the skyrmion simulation in a Poiseuille-like flow for different time steps for the finite-difference scheme $\Delta t _{FD}$, which are indicated on the top. All snapshots are for $t=1$s. The first column is our reference case, using double precision while the other columns are for single precision using different time steps. (a) and (b) depict the z-component of the director field while (d) and (c) depict the magnitude of the velocity field. In (c), $u_{max}=42\, \mu$m/s and $u_{min}=34\, \mu$m/s while, in (d), $u_{max}=41\, \mu$m/s and $u_{min}=0\, \mu$m/s. The white arrow in (a) indicate the main flow direction. }
\label{fig-screenshots}
\end{figure*}

\begin{table}
\caption{\label{tab1} Parameters used in the simulation and physical units.}
\footnotesize
\begin{tabular}{|p{0.28\linewidth}|p{0.28\linewidth}|p{0.28\linewidth}|}
\hline
symbol&sim. units & physical units\\
\hline
$\rho$&1&1088 Kg/m$^{3}$\\
\hline
$\Delta x$&1 & 0.625 $\mu$m\\
\hline
$\Delta t$&1 & 2$\times10^{-9}$ s\\
\hline
$K_{11}$&$1.67 \times 10^{-7}$& $6.4\times 10^{-12}$ N \\
\hline
$K_{22}$&$7.88 \times 10^{-8}$& $3.0\times 10^{-12}$ N \\
\hline
$K_{33}$&$2.62 \times 10^{-7}$& $9.98\times 10^{-12}$ N \\
\hline
$\alpha_1$& 0.0373 & 0.0036 Pa.s\\
\hline
$\alpha_2$& -0.4496 & -0.044 Pa.s\\
\hline
$\alpha_3$& -0.0203 & -0.0020 Pa.s\\
\hline
$\alpha_4$& 0.9318 & 0.091 Pa.s\\
\hline
$\alpha_5$& 0.3084 & 0.030 Pa.s\\
\hline
$\alpha_6$& -0.1617 & -0.016 Pa.s\\
\hline
$P$ & 14 & 8.75 $\mu$m \\
\hline
$L_x$, $L_y$, $L_z$& 56, 56, 16 & 35, 35, 10 $\mu$m\\
\hline
\end{tabular}\\
\end{table}
\normalsize

\subsection{Hybrid method}
\label{hybrid-sec}

The LBM and FD hybrid method can be implemented following the steps below, the sequence of which is illustrated in Fig.~\ref{fluxogram-fig}.
\begin{itemize}
  \item \textit{Initialization}. Initialize all the fields in LBM and FD, including the density, velocity, distribution functions and director field.
\item \textit{Equilibrium distribution function}. Use the velocity and density fields in Eq.~\eqref{shifted-feq}.
\item \textit{Collision and streaming}. Apply Eq.~\eqref{shifted-lbe}. Note that the force from Eq.~\eqref{force-lc-eq} is fixed during the same time step of the finite-differences $\Delta t_{FD}$.  
\item \textit{Macroscopic quantities}. Calculate the density and velocity fields using Eqs.~\eqref{shifted-rho} and~\eqref{shifeted-u}. Again, the force is fixed within a time step $\Delta t_{FD}$.
\item If the time is multiple of $\Delta t_{FD}$ ( $t\, \% \,\Delta t_{FD}=0$ ), then move the the FD method. Otherwise, go to the next LBM step.
\item \textit{Predictor-corrector}. Apply the predictor-corrector method described in Sec.~\ref{FD-sec}
\item \textit{Calculate the force}. Update the force using Eq.~\eqref{force-lc-eq}.
\item Continue to the first LBM step if the simulation has not finished or stop otherwise.
\end{itemize}

\subsection{Precision in GPU calculations}

Modern PC and workstation GPUs have substantial computing resources allocated for single-precision computations, much more than the resources allocated for double-precision. For instance, the workstation GPU used in this work (Nvidia RTX 4000 Ada) has 48 streaming multiprocessors (SMs), each SM is equipped with 128 CUDA cores that could be used in single precision (FP32) computations, while each SM is also equipped with only 2 CUDA cores that could be used in double precision (FP64) computations. Hence, the FP64 computations rate in Tera floating point operations (TFLOP) is 1/64th of the TFLOP rate of FP32 operations. The small number of FP64 Cores is included only to ensure that programs with FP64 code operate correctly on the GPU.
Since LBM is a memory-bound computational method, it is preferred to work with single-precision to optimize the use of the GPU memory bandwidth. The (float) datatype needs 4 bytes to store each number, while the (double) datatype needs 8 bytes to store each number. Hence, the GPU memory bandwidth could be used to transfer twice the amount of (float) numbers than the amount of (double) numbers. This has a major impact on the computational speed of the LBM.
In previous work, liquid crystal simulations were only possible in double precision as the single precision computations resulted in non-physical results. The small changes in the velocity field were not captured properly with single precision accuracy. That resulted in a major slowdown in the computational speed of the simulations due to the limited double-precision computational resources in the available GPUs. Therefore, the simulations were carried out for relatively small domains and a limited number of cases to obtain results in a timely manner. To alleviate those limitations, single precision computations were necessary to better utilize the available computational resources. However, the simulation accuracy was of great concern due to the limits of single-precision computations. In this work, we utilize the distribution functions shifting technique reported in Ref.~\cite{computation4010011} to increase the computational accuracy of the LBM, as shown in Sec.~\ref{lbm-sec}.

\section{Results}
\label{results-sec}

In this section, we present the results of the liquid crystal simulations using the shifting technique and the larger time step for the finite-difference scheme, as described in the previous section. We first analyze the results obtained using double precision, followed by those from single precision calculations.

\subsection{Double precision}
As a test case, we simulate a liquid crystal skyrmion in a Poiseuille flow, similar to those in our previous studies~\cite{Amaral2025, mi15111302, coelho2024halltransportliquidcrystal}. The system dimensions are $64 \times 64 \times 16$, with material parameters listed in Table~\ref{tab1}. The fluid is driven by an external force of $g=2\times 10^{-9}$ in all simulations. Periodic boundary conditions are applied in the x and y directions, while no-slip (half-way bounce-back) boundary conditions are enforced at the top and bottom solid plates.

For the director field, no anchoring is applied at the plates, allowing the formation of a ``skyrmionic tube''. To stabilize this configuration in the absence of anchoring, we introduce a weak effective anchoring throughout the domain with strength $W_0=0.000000015$, mimicking the high-frequency electric field used in experiments to stabilize skyrmions~\cite{Ackerman2017}. This trick was also used in our previous 2D simulations~\cite{PhysRevResearch.5.033210, Coelho_2021}. The director field is initialized at rest, following the ansatz from Ref.~\cite{Coelho_2021}. Simulations are run until $t=1$ s, corresponding to $5\times 10^{8}$ LBM iterations (or $\Delta t$).

\begin{figure*}[htb] \includegraphics[width=\textwidth]{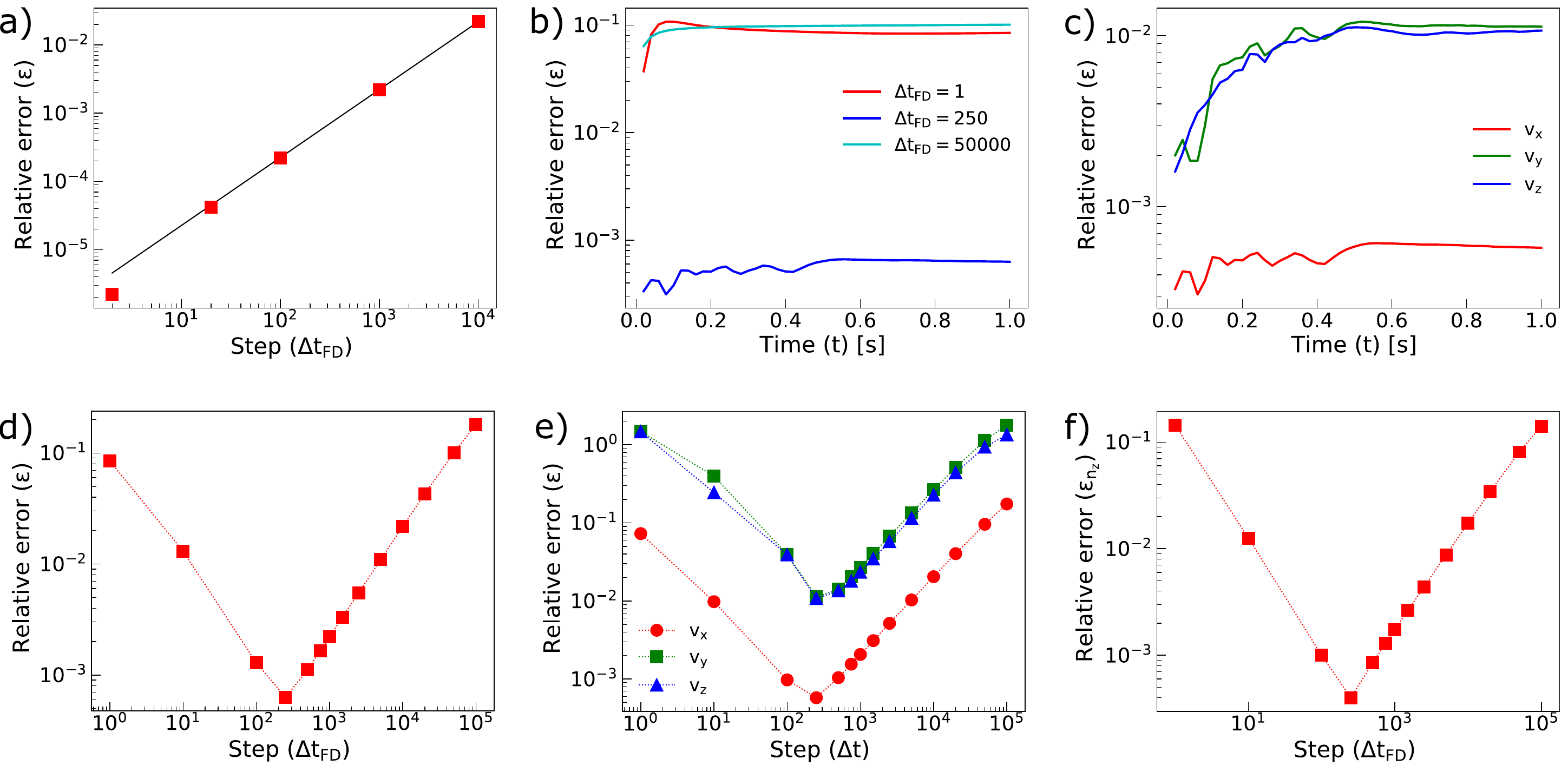} \caption{Relative error calculated at the end of the simulations (at $t=1$s) (a) Relative error $\varepsilon$ using double precision with different time steps $\Delta t_{FD}$ for the finite difference scheme. The solid line represents a power-law fit, with a slope of $1.07 \pm 0.04$. Time evolution of the relative error using single precision (b) for different time steps and (c) for the different velocity components using the optimal time step $\Delta t_{FD}=250$.
Relative error as a function of the time step for (d) the velocity field, (e) each of the three velocity components, and (f) the director field.} 
\label{precision-fig} 
\end{figure*}

When using double precision, the shifting technique produces results that are essentially identical. We consider the double-precision simulation with the shifting technique and $\Delta t_{FD}=1$ as the most accurate, treating it as our reference case. The relative error with respect to this reference is defined as:
\begin{eqnarray} \varepsilon = \frac{\sum _n \sqrt{ (u_x-u_x^{\text{ref}})^2 + (u_y-u_y^{\text{ref}})^2 + (u_z-u_z^{\text{ref}})^2}}{\sum _n \sqrt{ (u_x^{\text{ref}})^2 + (u_y^{\text{ref}})^2 + (u_z^{\text{ref}})^2}}, 
\end{eqnarray}
where the sum runs over all spatial points $n$. 
The relative difference for the case using double precision without the shifting technique is $\varepsilon=2.68\times 10^{-9}$, while the computational performance, measured in MLUPS (mega lattice updates per second), remains nearly the same: 78.68 MLUPS without the shifting technique and 76.70 MLUPS with it.

Figure~\ref{fig-screenshots}a presents snapshots of the director and velocity fields for our reference simulation, which will later be compared with single-precision results. Initially, the skyrmion has a symmetric configuration but evolves into a ``distorted cylinder'' in the steady state. The velocity field is nearly a Poiseuille flow, with minor perturbations near the skyrmion.

We plot histograms of the velocity and distribution functions for the double-precision simulation with $\Delta t_{FD}=1$. Figure~\ref{histograms-fig}a shows the velocity distribution, where no significant differences are observed between simulations with and without the shifting technique (not shown). Peaks at lower velocities arise due to spatial discretization along the z-direction: each layer has a well-defined velocity following the parabolic profile, but variations due to the skyrmion and liquid crystal structure lead to a broader velocity distribution at larger values.

A key observation is the order of magnitude of the velocity field, which reaches only $\sim 10^{-7}$ in lattice units. This is much smaller than in typical LBM simulations of Newtonian fluids using double precision to correctly store powers of $\vert \mathbf{u} \vert$ in the usual distribution function, Eq.\eqref{feq-eq}. This is because, in LBM, the distribution function is centred around the discrete weights of the Gauss-Hermite quadrature (D3Q19 in our case), as illustrated in Fig.\ref{histograms-fig}b. A clever approach proposed in Ref.\cite{computation4010011} involves shifting the distribution function so that it is centered around zero, as described in Sec.\ref{lbm-sec}. As demonstrated in Fig.~\ref{histograms-fig}c, the shifted distributions remain close to zero, enabling more efficient storage. This simple technique significantly enhances numerical accuracy and partially allows the use of single precision, which makes simulations on gaming GPUs much more efficient. But this improvement alone is not enough to run the full simulation in single precision because the director updates in each LBM time step are too small leading to additional precision problems.

Next, we examine the impact of different time steps $\Delta t _ {FD}$ in double-precision simulations, which serves as a reference for comparison with single precision. Figure~\ref{precision-fig} shows that the error $\varepsilon$ increases monotonically with the time step. A power-law fit yields a slope slightly above one ($1.07$). While the predictor-corrector scheme typically exhibits second-order convergence, this hybrid method is constrained by the need to compute the time derivative using only the velocity field at the current step, leading to a convergence rate below two. Additionally, the slope at smaller time steps is significantly larger than one, because assuming a constant velocity field between the current and next time step is a better approximation at lower $\Delta t_{FD}$.

\subsection{Single precision}

We now explore the feasibility of using single precision for the simulations with the method described in this paper. Figure~\ref{fig-screenshots} presents the results for the reference simulation (double precision, $\Delta t_{FD}=1$) alongside three different time steps using single precision. It is evident that for $\Delta t_{FD}=1$ and $\Delta t_{FD}=50000$, the results deviate visibly from the reference case, whereas for $\Delta t_{FD}=250$, there is no significant difference.

In Fig.~\ref{precision-fig}b, we plot the time evolution of the error for these three time steps using single precision. All curves stabilize around $t = 0.5$ s, indicating that running simulations up to $t = 1$ s is sufficient for error analysis. As suggested by the visual comparison, the error for $\Delta t_{FD}=250$ is significantly smaller than for the other two cases. This result implies that the error exhibits a non-monotonic dependence on the time step, differing from the behavior observed in double-precision simulations.

Figure~\ref{precision-fig}c presents the time evolution of the error for $\Delta t_{FD}=250$ for each velocity component $\alpha$:
\begin{eqnarray} \varepsilon = \frac{\sum_n \sqrt{ (u_\alpha-u_\alpha^{\text{ref}})^2} }{\sum_n \sqrt{ (u_\alpha^{\text{ref}})^2 }}. 
\end{eqnarray}
It can be observed that the error for the x-component is smaller than for the other two components. This is due to the larger absolute magnitude of the x-component, which reduces precision errors in the calculations.

\begin{figure}[htb] \includegraphics[width=0.45\textwidth]{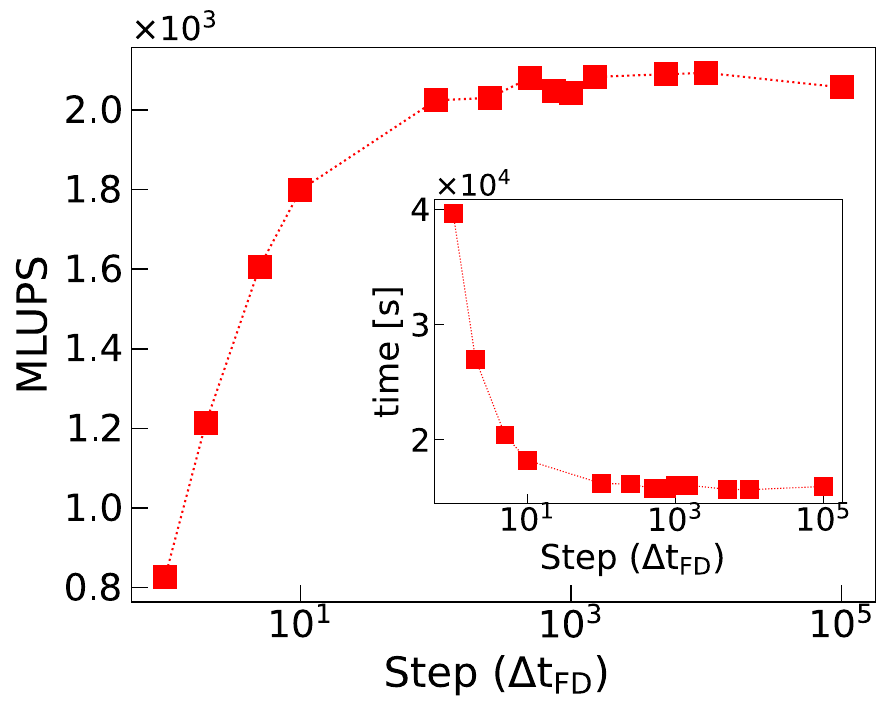} \caption{Simulation speed in mega lattice updates per second (MLUPS) as a function of the time step for single-precision simulations. The inset shows the total simulation time for each case.} 
\label{mlups-fig} 
\end{figure}

Figure~\ref{precision-fig}d reveals a non-monotonic relationship between the error and the time step, in contrast to the strictly increasing error observed in double-precision simulations. This behavior can be explained as follows: for small $\Delta t_{FD}$, the time derivative in Eq.~\eqref{director-time-eq} is too small, leading to precision errors. As $\Delta t_{FD}$ increases, the error initially decreases until reaching a minimum. Beyond this point, the error starts increasing again due to the intrinsic numerical error of the method, similar to the trend observed in double-precision simulations.

Therefore, choosing the optimal time step, $\Delta t_{FD}=250$, maximizes accuracy. At this value, the error reaches a remarkably low value of 0.063\%. This demonstrates that single-precision calculations can produce results nearly identical to those obtained with double-precision. This is particularly beneficial for GPU-based simulations, where most CUDA cores are optimized for single-precision arithmetic, significantly accelerating computations.

We also examine the error for individual velocity components, shown in Fig.~\ref{precision-fig}e. As previously noted, the x-component exhibits a smaller error due to its larger absolute value, which mitigates precision issues. Additionally, the error associated with the director field reaches a minimum at the same time step.

Another crucial factor in selecting the time step is computational performance. Figure~\ref{mlups-fig} shows the MLUPS as a function of the time step. The performance increases rapidly for small time steps and stabilizes around $\Delta t_{FD} \sim 100$. The same stabilization is observed for the total simulation time (inset). Since $\Delta t_{FD}=250$ lies beyond this stabilization region, it is a safe choice in terms of both accuracy and computational efficiency. Using $\Delta t_{FD}=250$ makes the simulation approximately 2.5 times faster compared to $\Delta t_{FD}=1$, while maintaining numerical reliability.

Although the same GPU was used for all simulations, the fluctuations observed in Fig.~\ref{mlups-fig} are attributed to uncontrollable hardware conditions, such as temperature variations, memory traffic to RAM, and output writing. The observed stabilization occurs because, at larger time steps, the computational cost of the LBM simulation surpasses that of the finite-difference scheme.

\begin{figure}[htb] \includegraphics[width=0.49\textwidth]{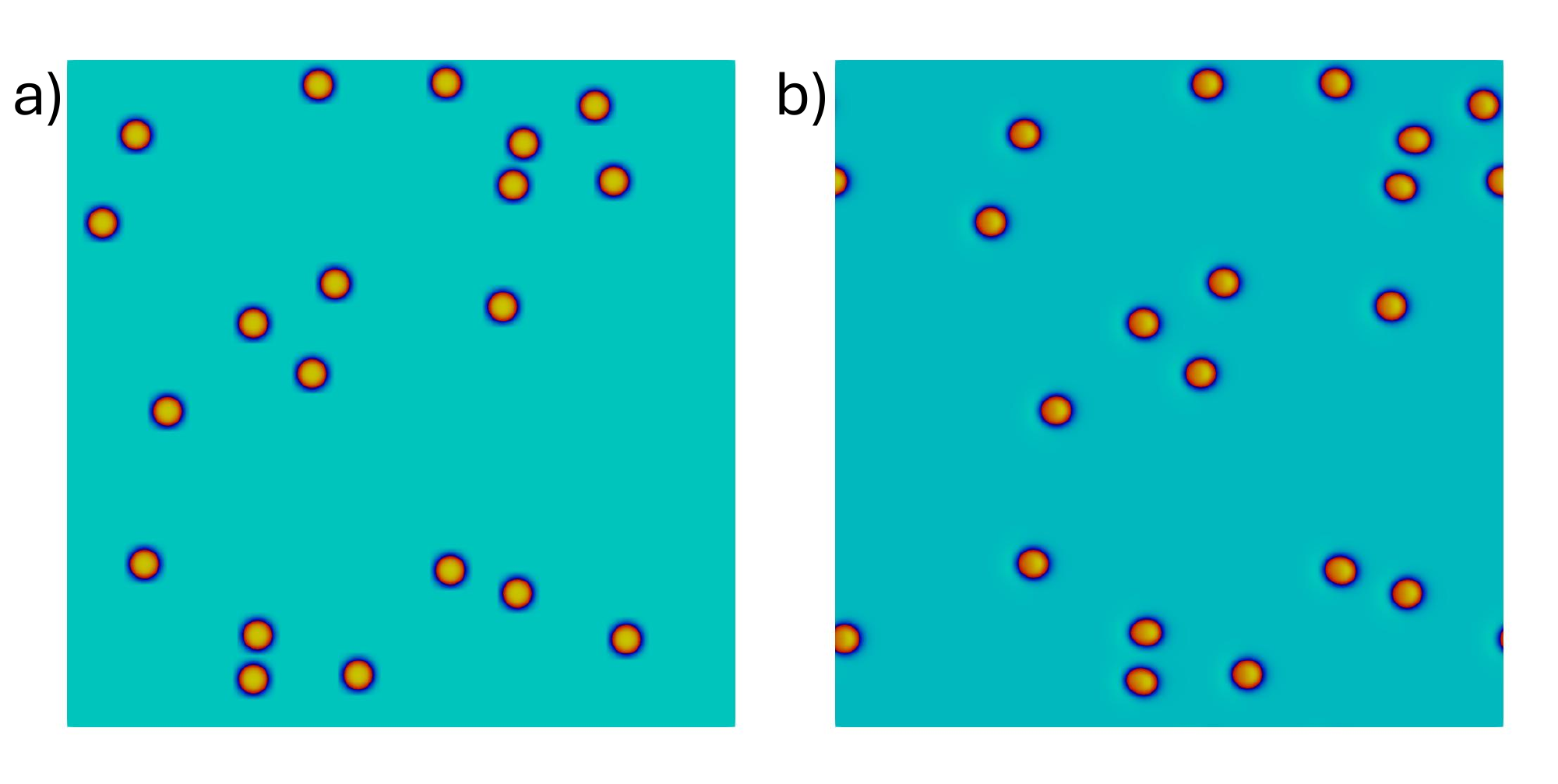} \caption{Snapshots of a system with many skyrmions (20) in a larger system with dimensions $320 \times 320 \times 16 \ $. (a) Initial configuration. (b) System at $t=1$s.} 
\label{many-fig} 
\end{figure}

To illustrate the capabilities of the method, we performed a simulation in a larger domain of $320 \times 320 \times 16 \ $ with 20 randomly placed skyrmions, as shown in Fig.~\ref{many-fig}. We used the model with single-precision and with $\Delta t_{FD}=250$. Similar simulations could be used in future studies to investigate collective effects between skyrmions or systems with higher resolution. Although the system is larger than those in previous studies, it ran within a perfectly feasible time on a gaming GPU (approximately five days).

\section{Conclusions}
\label{conclusion-sec}

We have presented an optimized algorithm based on the Ericksen-Leslie model to simulate the hydrodynamics of liquid crystals. Our method combines the lattice Boltzmann method (for the velocity field) with finite differences (for the director field). We introduce two key improvements to enhance precision and enable efficient single-precision computations.

The first improvement is the shifted distribution function in the lattice Boltzmann method. Since typical velocities in these simulations are extremely low ($\sim 10^{-7}$), standard lattice Boltzmann implementations suffer significant precision loss. The shifting technique mitigates this issue, substantially improving accuracy. However, this enhancement alone is insufficient for reliable single-precision simulations because the director field updates are too small in each lattice Boltzmann time step, leading to further precision errors.

The second improvement is decoupling the time steps of the lattice Boltzmann and finite-difference methods. By using larger time steps for finite differences, we reduce precision errors and enable the entire simulation to run in single precision without sacrificing accuracy.

Our results show that, in double precision, accuracy decreases monotonically as the finite-difference time step increases. However, in single precision, accuracy follows a non-monotonic trend: initially, the error decreases as the time step increases due to larger director field updates, reaches a minimum, and then grows due to method-induced errors. This behaviour reveals an optimal time step that maximizes accuracy. In our test case, the error was remarkably low ($\approx 0.06$\%) compared to double-precision simulations, with a speedup of $26.5$ times. We illustrate the possibilities opened by this technique by simulations of a large domain with many flowing skyrmions.

Single precision is crucial for GPU-accelerated simulations, as consumer-grade gaming GPUs, which are significantly cheaper and more widely available than scientific GPUs, are optimized for single-precision calculations. Moreover, the techniques introduced here can be extended to other hybrid lattice Boltzmann and finite-difference methods, broadening their applicability in computational fluid dynamics.

\section*{Acknowledgements}
We acknowledge financial support from the Portuguese Foundation for Science and Technology (FCT) under the contracts: PTDC/FISMAC/5689/2020 (DOI 10.54499/PTDC/FIS-MAC/5689/2020), UIDB/00618/2020 (DOI 10.54499/UIDB/00618/2020), UIDP/00618/2020 (DOI 10.54499/UIDP/00618/2020), DL 57/2016/CP1479/CT0057 (DOI 10.54499/DL57/2016/CP1479/CT0057) and 2023.10412.CPCA.A2 (DOI 10.54499/2023.10412.CPCA.A2).

\bibliography{biblio}

\end{document}